\documentclass[10pt,preprint]{aastex}
\usepackage{epsfig}

\def\simge{%
    \mathrel{\rlap{\raise 0.511ex
        \hbox{$>$}}{\lower 0.511ex \hbox{$\sim$}}}}
\def\simle{%
    \mathrel{\rlap{\raise 0.511ex
        \hbox{$<$}}{\lower 0.511ex \hbox{$\sim$}}}}

\begin{document}
\title{CONSTRAINING THE EQUATION OF STATE WITH MOMENT OF
INERTIA MEASUREMENTS}
\author{James M. Lattimer}
\affil{Department of Physics and Astronomy, Stony Brook University, 
Stony Brook, NY-11794-3800, USA}
\email{lattimer@mail.astro.sunysb.edu}
\author{Bernard F. Schutz}
\affil{Max-Planck Institut f\"ur Gravitationsphysik 
(Albert-Einstein-Institut),
Am M\"uhlenberg 1, D-14476, Potsdam, Germany}
\email{Bernard.Schutz@aei.mpg.de}
\shorttitle{Constraining the Equation of State}
\shortauthors{LATTIMER \& SCHUTZ}

\begin{abstract}
We estimate that the moment of inertia of star A in the recently discovered 
double pulsar system PSR J0737-3039 may be determined after a few years of
observation to something like 10\% accuracy. This would 
enable accurate estimates of the radius of the star and the presure of
matter in the vicinity of 1 to 2 times the nuclear saturation density, 
which would in turn provide strong constraints on the equation of state 
of neutron stars and the physics of their interiors. \\

\end{abstract}

\section{INTRODUCTION}

The discovery of the double-pulsar system PSR J0737-3039~\citep{Burgay03,
Lyne04} provides physicists with a remarkable laboratory for relativistic
astrophysics. Besides its implications for the rate of gravitational wave
bursts from neutron-star coalescence~\citep{Burgay03} and for the
understanding of pulsar magnetospheres~\citep{Lyne04}, it could provide a
measurements of spin-orbit coupling. \cite{Lyne04} noted that accurate timing
over a period of years could lead to a determination of the moment of inertia
of star A. Spin-orbit coupling could be revealed either through an extra
advancement of the periastron of the orbit above the standard post-Newtonian
advance, or in the precession of the orbital plane about the direction of the
total angular momentum of the system. Given that the masses of both stars are
already accurately determined by observations, a measurement of the moment of
inertia of even one neutron star could have enormous importance for neutron
star physics.

Despite the fact that over a thousand neutron stars have been discovered in
radio and X-ray observations, and accurate masses have been determined for a
dozen or so neutron stars in radio pulsar binaries~\citep{Stairs04a}, there is
relatively little observational information about their radii or their
interior physics. Currently, one source of candidates for revealing radii are
those neutron stars for which thermal x-ray and optical radiation have been
observed (see~\cite{Page04} for a review). Nevertheless, these stars have not
yet provided a clear radius determination, because of ambiguities due to
atmospheric re-processing, interstellar absorption, and distances. Moreover,
the inferred radius actually refers to the radiation radius,
$R_\infty=R/\sqrt{1-2GM/Rc^2}$, which is mass-dependent; unless independent
mass measurements of the same stars become available, such as would be
provided by a redshift, the radii themselves remain unknown. However, 
most of these sources have featureless spectra and no secure redshifts have
been obtained to date from them.

Another possibility are X-ray bursts from the surfaces of neutron stars.
Recently, two lines observed in an X-ray burst spectrum have been suggested to
be H- and He-like Fe lines and imply a redshift of 0.35 \citep{Cottam02}. This
inference has been given additional credibility by the detection
\citep{Villarreal04} of a 45 Hz rotational frequency for the neutron star, EXO
0748-676. A low spin rate is consistent with the observed equivalent widths of
these lines if their identifications with Fe are correct. \cite{Villarreal04}
show that this consistency holds if the neutron star radius is in the range
9.5--15 km (corresponding to masses in the range 1.5--2.3 M$_\odot$). Since
this star is a member of an eclipsing binary, an independent mass measurement
might be possible, as well. These techniques hold promise of further
constraining the radii of this star, and could be extended to other X-ray
bursters if redshifts from them can be observed.  

At present, however, the constraints on the neutron star radius from X-ray
bursters and thermal radiation from cooling neutron stars are relatively weak.
We demonstrate that securing a value for the moment of inertia for a star in a
radio binary pulsar system will provide important constraints on the radius of
the star and the equation of state of neutron-star matter.  Dimensionally, 
the moment of inertia is proportional to the star's mass times its radius 
squared, so that a measurement of the moment of inertia to a given accuracy
provides approximately twice the accuracy for a radius identification.

We begin in \S 2 by 
exploring the remark in~\cite{Lyne04} that the moment of inertia can
be determined from measurements of spin-orbit coupling. Two observable
effects pf spin-orbit coupling, precession of the orbital plane and an
extra contribution to the advance of the periastron, are investigated.
In \S 3, the case of PSR 0737-3039 is investigated.  We conclude that
a 10\% measurement of the moment of inertia of star A in this system
might be possible with observations extending over a period of a few
years. In \S 4 section, we show how such a measurement leads to an
estimate of the radius of the star and the density of neutron star
matter in the vicinity of the nuclear saturation density, This could
have crucial implications for delimiting the equation of state.  The
conclusion section, \S 5, compares the PSR0737-3039 system with other
relativistic binaries, and possibilities for the future of this
technique are discussed.

\section{OBSERVABLE SPIN-ORBIT EFFECTS}

There are two kinds of spin-coupled precession effects in a binary
system of compact stars: spin-orbit and spin-spin
couplings. Spin-orbit coupling leads to a precession of the angular
momentum vector $\vec{L}$ of the orbital plane around the direction of
the total angular momentum $\vec{J}$ of the system. This is sometimes
called geodetic precession, and is related to the Thomas precession of
atomic physics. Since the total angular momentum $\vec J=\vec L+\vec
S_A+\vec S_B$ is conserved (at this order), there are compensating
precessions of the spins $\vec{S}_A$ and $\vec{S}_B$ of the two
stars. Since the orbital angular momentum dominates the spin angular
momenta in binaries, the geodetic precession amplitude is very small
while the associated spin precession amplitudes are substantial.  In
addition to geodetic precession, spin-orbit coupling also manifests
itself in apsidal motion (advance of the periastron).  Spin-spin
coupling is negligible in close binary systems because $|\vec
L|>>|\vec S_A|,|\vec S_B|$.

According to \cite{Barker75}, 
the spin and orbital angular momenta evolve according to
\begin{eqnarray}
\dot{\vec S_i}&=&{G(4M_i+3M_{-i})\over2M_i a^3c^2(1-e^2)^{3/2}}
\vec L\times\vec S_i,\label{eqn:evolve,a}\\
\dot{\vec L^{SO}}&=&\sum_i{G(4M_i+3M_{-i})\over2M_ia^3c^2(1-e^2)^{3/2}}
(\vec{S_i}-3{\vec L\cdot\vec{S_i}\over|\vec L|^2}\vec L)\label{eqn:evolve,b}
\end{eqnarray}
where the superscript $SO$ refers to the spin-coupling contribution
only (there are also first- and second-order post-Newtonian terms, 1PN
and 2PN, respectively, unrelated to the spins, that contribute to this
order).  Here $a$ is the semimajor axis of the effective one-body
orbital problem (sum of the semi-major axes of the two stellar
orbits), $e$ is the eccentricity, and $M_i$ and $M_{-i}$ refer to the
masses of the two binary components (we use the notation that $i=A,B$
and $-i=B,A$).  To this order, one may employ the Newtonian relation
for the orbital angular momentum:
\begin{equation}
|\vec L|={2\pi\over P} {M_AM_Ba^2(1-e^2)^{1/2}\over 
M}=M_AM_B\sqrt{Ga(1-e^2)\over M}\,
\end{equation}
where $P$ is the orbital period and $M+M_A+M_B$.  
Then, from Eq. \ref{eqn:evolve,a}, 
the spin precession periods are
\begin{equation}\label{eqn:kidder} 
P_{p,i} = \frac{2c^2aPM(1-e^2)}{GM_{-i}(4M_i+3M_{-i})}\,, 
\end{equation} 
which are not identical for the two components unless they are of
equal mass.  Note that the spin precession periods are independent of the
spins.  Also note that if the spins are parallel to $\vec L$, there is,
first, no spin precession, and second, the spin-orbit contribution
to the advance of the periastron is in the sense opposite to the
direction of motion.

The spin precession leads to two observable effects.  First, as the spin
axis change orientation in space, the pulsar beams will sweep through 
changing directions in space.  In many cases, this will lead to the
periodic appearance and disappearance of the pulsar beam from the Earth.  
Second, since total angular momentum is conserved (to this order), the
orbital plane will change orientation.  This will be observed as a change
in the inclination angle $i$.

\cite{Damour88} have considered the question of how these effects affect
the timing of binary pulsars.  For the change in inclination, they find
\begin{eqnarray}\label{eqn:idot}
{di\over dt}={G\over ac^2}{\pi\over (1-e^2)^{3/2}}
\sum_i{I_i(4M_i+3M_{-i})\over M_ia^2P_i}\sin\theta_i\cos\phi_i\,,
\end{eqnarray}  
where $\theta_i$ is the angle between $\vec{S_i}$ and $\vec{L}$ and
$\phi_i$ is the angle between the line of sight to pulsar $i$ and the
projection of $\vec{S_i}$ on the orbital plane.  These angles follow the
convention of \cite{Jenet04}, but other references employ $\phi_i-90$ in
place of $\phi_i$, {\it cf.}, \cite{Wex99}.  Also, we used $|\vec{S_i}|=2\pi
I_i/P_i$ where $I_i$ and $P_i$ are the moment of inertia and the spin
period, respectively, of component $i$.  If we specialize to the case
in which the spin of one component is much greater than that of the
other, $|\vec{S_A}|>>|\vec{S_B}|$, the amplitude of the precession of
the inclination of the orbital plane is given by the change in
$\vec{L}$ needed to compensate changes in $\vec{S_A}$.  We define the
angle between $\vec L$ and $\vec J$ to be $\theta-\lambda$ and the
angle between $\vec {S_A}$ and $\vec J$ to be $\lambda$.  Since
$|\vec{S_A}|<<|\vec{L}|$, one has $|\theta-\lambda|<<|\lambda|$.
Using the fact that $\vec J\simeq\vec L+\vec {S_A}$, one finds $|\vec
L|\sin(\theta-\lambda)\simeq|\vec{S_A}|\sin\lambda\simeq|\vec{S_A}|\sin\theta$.
Thus, the amplitude of the change in the orbital inclination angle $i$
due to A's precession will be
\begin{equation}\label{eqn:i}
\delta_i={|\vec{S_A}|\over|\vec L|}\sin\theta_A 
\simeq{I_AM\over a^2M_AM_B(1-e^2)^{1/2}}{P\over P_A}\sin\theta_A\,.
\end{equation}  
This will cause a periodic departure from the expected
time-of-arrival of pulses from pulsar A of amplitude  
\begin{equation}\label{eqn:deltai}
\delta t_A={M_B\over M}{a\over c}\delta_i\cos{i}=
{a\over c}{I_A\over M_Aa^2}{P\over P_A}\sin\theta_A\cos i\,,
\end{equation}
if one can assume the orbital eccentricity is small.

For the advance of the periastron, the ratio of the spin-orbit
and 1PN contributions is \citep{Damour88}
\begin{eqnarray}\label{eqn:periastron}
{A_p\over A_{1PN}}=-{P\over6(1-e^2)^{1/2}Ma^2}
\sum_i{I_i(4M_i+3M_{-i})\over M_iP_i}
(2\cos\theta_i+\cot i\sin\theta_i\sin\phi_i)
\end{eqnarray}
In the case that $|\vec{S_A}|>>|\vec{S_B}|$, only the $i=A$ term contributes
substantially. For comparison, the ratio of the 2PN to 1PN contributions is
\citep{Damour88}
\begin{eqnarray}\label{eqn:a2pn}
{A_{2PN}\over A_{1PN}}&=&{GM\over 4ac^2}\sum_i\left
(\left[27+6{M_i\over M}+6\left({M_i\over M}\right)^2\right](1-e^2)^{-1}-1-
{46M_i\over3M}+{10\over3}\left({M_i\over M}\right)^2\right)\nonumber\\
&\simeq&{GM\over24ac^2}\left({189\over1-e^2}-47\right)\,,
\end{eqnarray}
where both $i=A$ and $i=B$ terms contribute. The second line of Eq.
\ref{eqn:a2pn} is valid in the case that $M_A=M_B$.

In \S4 we demonstrate that a very useful constraint on the equation of state
(EOS) can be made if the moment of inertia can be determined to about 10\%. In
practice, it is expected that the binary components will have approximately
equal masses but differing spin periods. Therefore, the spin-orbit effects
will be dominated by the more radidly rotating star, A. Besides being
functions of the parameters $M_A, M_B, P, a, e$ and $i$, the observables
$\delta t_A$ and $A_p$ are also functions of $\theta_A$ and $\phi_A$.
Therefore, extraction of $I_A$ from these observables requires that additional
information about the orientation of the spin of A be available. Fortunately,
if A is observed as a pulsar, observations of the beam geometry and its
precession can provide this information.
  
\section{APPLICATION TO PSR 0737-3039}

The observational parameters for the system PSR 0737-3039 are summarized in
\cite{Lyne04}: $M_A\simeq1.34{\rm~M}_\odot$, $M_B\simeq1.25{\rm~M}_\odot$,
$a/c = 2.93$ s, $e\simeq0.088$, $P_A\simeq22.7\rm{~ms}$, $P_B\simeq2.77 $ s,
and $P\simeq0.102$ day. We therefore observe that $P_{pA}\simeq74.9$ yrs and
$P_{pB}\simeq70.6$ yrs. With these parameters, we can form the useful
combinations
\begin{eqnarray}
{GM\over ac^2}=4.32~10^{-6}\,,\qquad 
{I_A\over Ma^2}=(7.74~10^{-11})~I_{A,80}\,,\qquad 
{P\over P_A}=3.88~10^5
\end{eqnarray}
where $I_{A,80}=I_A/(80 {\rm~M}_\odot{\rm~km}^2)$ is a typical value for
the moment of inertia (see \S 4). 

Since $P_B/P_A=122$, we can ignore contributions to the precession
from pulsar B. The orientation of the spin axis of $A$ relative to the
orbital plane has been estimated through modeling of the intensity
variations of pulsar B due to illumination by emission from A
\citep{Jenet04}. There are two solutions to this model: solution 1
with $\theta_A=13^\circ\pm10^\circ$ (or $167^\circ\pm10^\circ$) and
$\phi_A=246^\circ \pm5^\circ$, and solution 2 with
$\theta_A=\pm90^\circ\pm10^\circ$ and $\phi_A=239^\circ\pm2^\circ$.
However, solution 1 is preferred on two grounds: it is improbable that
a supernova kick would result in a spin axis that is so strongly
misaligned with the orbital angular momentum, and solution 2 is
also inconsistent with the misalignment angle between $\vec{S_A}$ and
the magnetic dipole axis, measured to be a few degrees
\citep{Demorest04}.  We assume solution 1 for the following
discussion.  The orientation of B has not been estimated so far, but
this is irrelevant since the spin of pulsar B can be ignored.

Recent determinations of the inclination angle are
$i=88.4^{\circ}~^{+1.6^\circ}_{-1.4^\circ} $ \citep{Ransom04} and
$i=90.26^\circ\pm0.13^\circ$ \citep{Coles04}. We shall use the latter as being
the more precise. The facts that the orbit is seen nearly edge-on and that
$\vec{S_A}$ is only slightly misaligned from $\vec L$ makes this a special
case in which the amplitude of the timing change produced by the orbial plane
precession will be extremely small, $\delta
t_A\simeq(0.17\pm0.16)I_{A,80}~\mu$s. Not only is the magnitude very small,
but the large relative uncertainties in both $\cos i$ ($|\cos
i|\simeq0.0045\pm0.0023$) and in $\sin\theta_A$
($\sin\theta_A\simeq0.22\pm0.17$) combine to give a huge uncertainty in
$\delta t_A$. Current technology allows timing accuracies at the level of
tenths of microseconds, limited by the quality of clocks and the
characterization of radio antenna signal paths~\citep{Lorimer01}. Given this
sensitivity, observation of the changing inclination due to orbital precession
seems marginal. Even if timing sensitivities to order 0.01$\mu$s were
possible, errors in both $i$ and $\theta_A$ would have to be substantially
reduced in order to measure $I_A$ to order 10\%.

The periastron advance due to spin-orbit coupling, being proportional to $\sin
i$ and $\cos\theta_A$, is, on the other hand, less sensitive to errors in
these angles. In addition, since $i\simeq90^\circ$, Eq. \ref{eqn:periastron}
shows that uncertainties in the angle $\phi_A$ become largely irrelevant. The
1PN perihelion advance is
\begin{eqnarray}
A_{1PN}={6\pi\over1-e^2}{GM\over ac^2} {\rm~radians~per~orbit}\,,
\end{eqnarray}
or 0.294 radians per year.
The periastron advance ratio is
$A_{pA}/A_{1PN}\simeq6.6^{+0.2}_{-0.6}~10^{-5}I_{A,80}$.

In practice, to measure the spin-orbit effect to 10\% means subracting out the
spinless pieces of the perihelion precession ($A_{1PN}$ and $A_{2PN}$) to one
part in about $10^5$. Fortunately, both $a\sin i$ and $P$ are well known, and
will be even better known in the future, and the spinless periastron advance
depends only on the total mass, not the individual masses. The total mass
$M=a^3(2\pi/P)^2$ is known to the accuracy of $\sin i$. But $\sin i$ is
currently known to about 2 parts in a $10^6$, using the determination of
\cite{Coles04}, and its accuracy will also improve with time. So measurement
of the spin-orbit perihelion advance seems possible.

Nevertheless, the determination of $I_A$ from this technique will only be as
accurate as our understanding of $\theta_A$. An uncertainty in $\cos\theta$ of
better than 6\% has already been achieved. Further improvements in our
knowledge of the pulse geometry, which seems likely, suggest that a moment of
inertia determination accurate to 10\%, may be possible. The epoch of
periastron changes due to spin-orbit coupling by about 0.027 seconds per year.
For comparison, the second post-Newtonian correction contributes a periastron
advance time of about 0.02 seconds per year and must therefore be included in
the pulsar timing model.

\section{CONSTRAINING THE EOS}

\begin{figure}[htb]
\begin{center}
\includegraphics[scale=.5,angle=90]{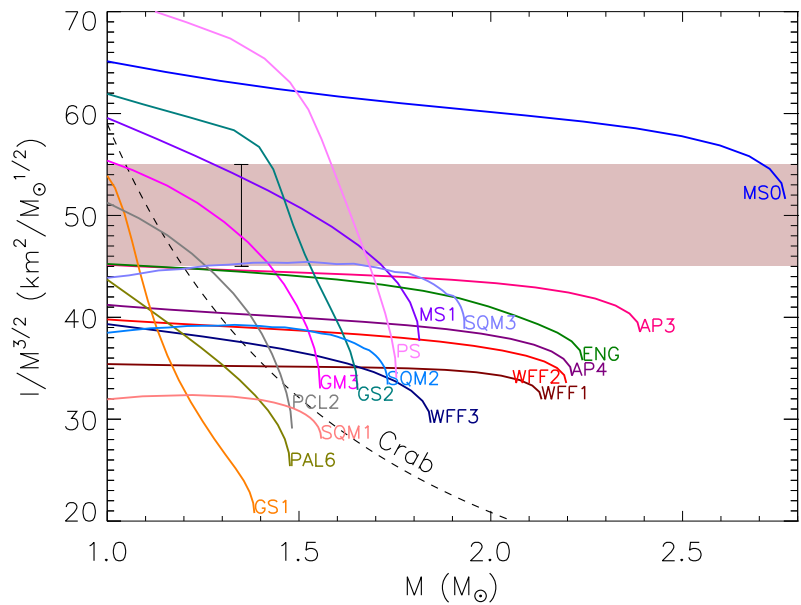}
\end{center}
\caption{The moment of inertia
scaled by $M^{3/2}$ as a function of stellar mass $M$.  Equation of
state labels are described in \cite{Lattimer01}.  The shaded
band illustrates a 10\% error on a hypothetical $I/M^{3/2}$ measurement
of 50 km$^2$ M$_\odot^{-1/2}$; the error bar shows the specific case in
which the mass is 1.34 M$_\odot$.  The dashed curve labelled "Crab" is
the lower limit derived by \cite{Bejger03} for the Crab pulsar.}
\label{mom}
\end{figure}

The estimation of a neutron star's moment of inertia from timing
observations of a radio binary pulsar has significant implications for
constraining the equation of state (EOS). In some respects, a moment
of inertia measurement could be more useful than a radius measurement
of the same accuracy. First, the neutron star mass in a radio binary
will obviously already be known to high precision, while a radius
measurement from observations of thermal radiation actually refers to
the radiation radius $R_\infty=R/\sqrt{1-2GM/Rc^2}$. To obtain the
radius and mass separately requires an estimate of the star's redshift
$(1-2GM/Rc^2)^{-1/2}-1$ which so far is not yet available for any
thermal source. Moreover, the accuracy of radiation radius estimates
from thermal sources are limited by uncertainties due to source
atmospheric re-processing, interstellar absorption, and distances. In
the case of X-ray bursters, the one case in which a redshift
measurement may be secure, EXO 0748-676, no independent mass
measurement is yet available, and radius information obtained from
fitting the line profile with the observed rotation rate is relatively
weak. Mass and moment of inertia measurements from radio binary
pulsars are not sensitive to estimates of the distance, and a redshift
is not required.  Second, the range of moments of inertia for various
neutron star models (including strange quark matter stars) is, in
relative terms, larger than the predicted range of radii (see
Fig. \ref{mom}). This follows from the dimensional relation $I\propto
MR^2$. For example, it is 30 to 240 M$_\odot$ km$^2$ for masses larger
than 1 M$_\odot$, and is 53 to 109 M$_\odot$ km$^2$ for a 1.4
M$_\odot$ star. For comparison, the range of radii for a 1.4 M$_\odot$
model is 9 to 16 km.

Spanning the same set of equations of state as in the compendium of
\cite{Lattimer01}, moments of inertia for normal neutron and strange quark
matter stars are displayed in Fig. \ref{mom}. The moments of inertia have been
scaled by a factor $M^{3/2}$ to reduce the range of the ordinate. For most
masses, the range in $I$ is approximately a factor of 2 to 3. The significance
of a measurement of $I$ with $\pm10$\% accuracy is illustrated by the shaded
band centered on the hypothetical measurement, here taken to be
$I/M^{3/2}=(50\pm5)$ km$^2$ M$^{-1/2}_\odot$, together with an error bar
located at a precisely measured mass, taken to be 1.34 M$_\odot$. It
is clear that relatively few equations of state would survive these
constraints. Those families of models lying close to the measured values would
have their parameters limited correspondingly.

\begin{figure}[htb]
\begin{center}
\includegraphics[scale=.5,angle=90]{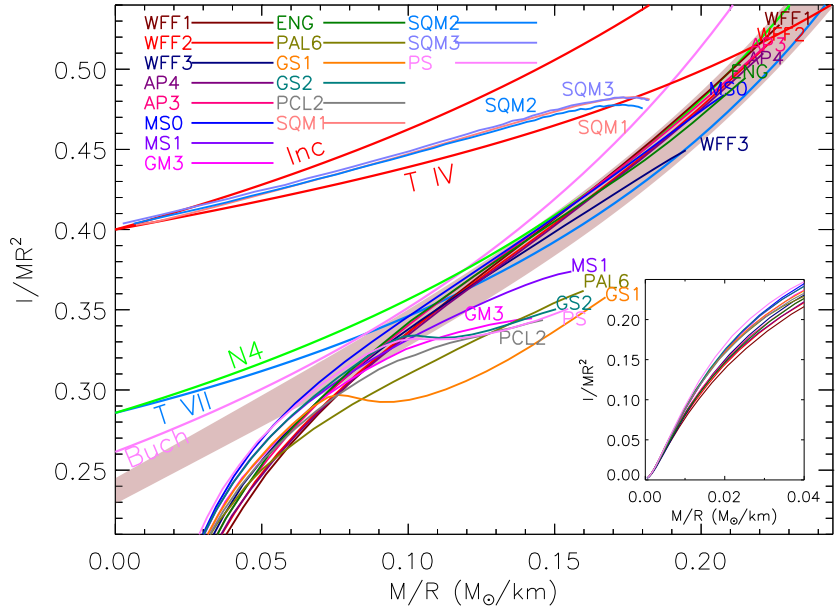}
\end{center}
\caption{The moment of inertia as a function of the relativity parameter
$M/R$.  The curves labelled
TIV, TVII, NIV, Inc and Buch refer to analytic solutions of Einstein's
equations, while other curves refer to a variety of equations of state (see
\cite{Lattimer01} for details). The shaded band illustrates the relation Eq.
\ref{eqn:imr2}.}
\label{mommr}
\end{figure}

If the equation of state does not have a large degree of softening at
supernuclear densities, possibly introduced by hyperons, Bose condensates or
self-bound strange quark matter, a moment of inertia determination furthermore
permits one to estimate the neutron star radius to a relative uncertainty
smaller than the relative uncertainty in the moment of inertia measurement.
Fig. \ref{mommr} shows the moment of inertia as a function of the relativity
parameter $M/R$ for the same equations of state displayed in Fig. \ref{mom}.
Several analytic solutions of Einstein's equations that are applicable either
to normal or self-bound stars. are displayed as well. These solutions are all
scale-free and are functions of $M/R$ alone; hence, they cannot be displayed
in Fig. \ref{mom}. Unless the equation of state has an appreciable degree of
softening, usually indicated by a maximum mass of order 1.6 M$_\odot$ or less,
or unless it is self-bound, there appears to be a relatively unique relation
between $I/MR^2$ and $M/R$. For $M/R$ values greater than approximately 0.07
M$_\odot$ km$^{-1}$, {\it i.e.}, for $M\ge1.0$ M$_\odot$, this relation can be
approximated by
\begin{equation}\label{eqn:imr2}
I\simeq (0.237\pm0.008)~MR^2~\left[1+4.2
{M{\rm~km}\over{\rm M}_\odot R}+90\left({M{\rm~km}\over{\rm M}_\odot
R}\right)^4\right].
\end{equation}
An analogous fit has also been suggested by \cite{Bejger02}:
\begin{equation}\label{eqn:bejger}
I\simeq{2\over9}\left[1+5{M{\rm~km}\over R{\rm~M}_\odot}\right]
\end{equation}
for $M/R>0.1{\rm~M}_\odot$/km; however, this fit underestimates $I$ for the
largest neutron star masses.

\begin{figure}[htb]
\begin{center}
\includegraphics[scale=.5,angle=90]{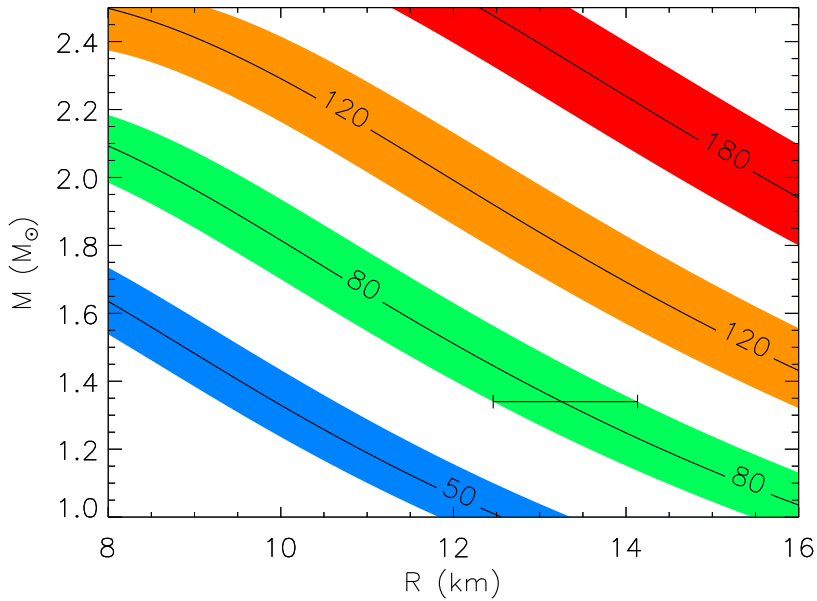}
\end{center}
\caption{Radius limits imposed by simultaneous moment of inertia and
mass measurements, established from Eq. \ref{eqn:imr2}.  Moment of
inertia error bands include measurement uncertainties of 10\% and
systematic uncertainties from Eq. \ref{eqn:imr2} and are labelled in
units of M$_\odot$ km$^2$.  The horizontal error bar illustrates the
hypothetical case in which $M$ and $I$ are measured to be 1.34
M$_\odot$ and $80\pm8$ M$_\odot$ km$^2$, respectively.}
\label{rcon}
\end{figure}

Simultaneous mass and moment of inertia measurements could therefore usefully
constrain the radius. The relevant radius relation can be determined by
inversion of Eq. \ref{eqn:imr2}. Fig. \ref{rcon} shows how the radius could be
constrained for selected moment of inertia measurements having 10\%
uncertainty. For a 1.4 M$_\odot$ star, this typically results in a radius
estimate with about 6--7\% uncertainty. Even in the event of significant
softening of the equation of state, the uncertainty of the estimated radius
would be degraded by no more than a factor of two. Of course, the accumulation
of more than a single $I-M$ pair would significantly enhance the constraints.

\begin{figure}[htb]
\begin{center}
\includegraphics[scale=.5,angle=90]{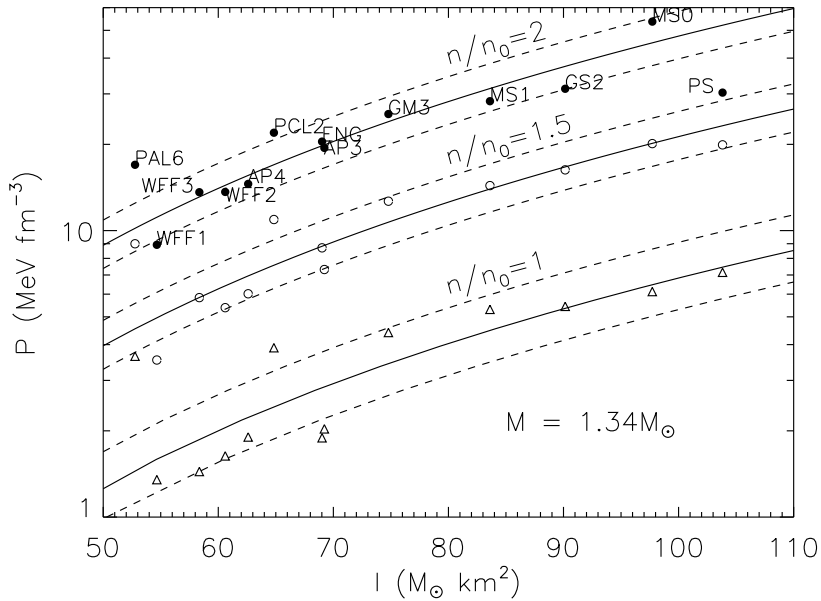}
\end{center}
\caption{Pressures at the densities $n_s$ (hollow triangles), $1.5n_s$
(hollow circles) and $2n_s$ (filled circles) as a function of the
moment of inertia for 1.34 M$_\odot$ stars. Solid curves show the
relations between $P$ and $I$ derived by combining Eqs.
(\ref{eqn:imr2}) and (\ref{eqn:rp}). Dashed curves include the errors
in these fits.  Equations of states and their labels are described in
\cite{Lattimer01}.}
\label{pi}
\end{figure}

The importance of a radius determination is that it immediately translates
into a measure of the neutron star matter pressure near the nuclear saturation
density \citep{Lattimer01}  In particular, the relation between these
quantities is of the form of a power law
\begin{equation}
R_MP(n)^{-1/4}=C(n,M).
\label{eqn:rp}
\end{equation}
Here $R_M$ is the neutron star radius in km for the mass $M$ and $P(n)$ is the
matter pressure in MeV fm$^{-3}$ evaluated at the density $n$. The constant
$C$ is parametrized by both $n$ and $M$, but its dependence upon $M$ is weak
(approximately scaling as $M^{-1/8})$. In the case of $M=1.4$ M$_\odot$, a
least squares fit to approximately 30 equations of state yielded
$C(n,1.4)=9.30\pm0.60, 7.00\pm0.31$ and $5.72\pm0.25$ for the cases $n/n_s=1,
1.5$ and 2, respectively. This relation could be made more precise by
adjusting the exponent of $P$ \citep{Steiner04}, but we choose not to do so
here. The pressure of neutron star matter at these densities is primarily a
function of the density dependence of the nuclear symmetry energy. In general,
therefore, we expect that measurement of the moment of inertia would provide
estimates of pressures. Fig. \ref{pi} illustrates the situation for some
representative equations of state for the case of $M=1.34$ M$_\odot$. It is
observed that the phenomenological fits Eqs. (\ref{eqn:imr2}) and
(\ref{eqn:rp}) adequately describe all but the softest equations of state
(e.g., PS and PAL6) for each density. The uncertainties in the estimated 
pressure obtained from a measured value of the moment of inertia are moderate,
amounting to a factor of about 2 when a 10\% uncertainty in the moment of
inertia measurement is included. Nevertheless, given the fact that present
estimates of the pressure of matter at the nuclear saturation density span a
range of a factor of 6, this information will be valuable.

\section{DISCUSSION AND CONCLUSIONS}

\begin{deluxetable}{lllll}
\tablecolumns{4}
\tablecaption{Comparison of Binary Pulsars
\label{tab:bp}}
\tablehead{
\colhead{}&
\colhead{PSR 0707-3039}\qquad &
\colhead{PSR 1913+16}\qquad &
\colhead{PSR 1534+12}\qquad &
}
\startdata
References & a, b, c & d, e & f, g, h \\

$a/c$ (s) & 2.93 & 6.38 & 7.62 \\

$P$ (h) & 2.45 & 7.75 & 10.1 \\

$e$ & 0.088 & 0.617 & 0.274 \\

$M_A$ (M$_\odot$)& 1.34 & 1.441 & 1.333 \\

$M_B$ (M$_\odot$) & 1.25  & 1.387 & 1.345  \\

$T_{GW}$ (Myr) & 85 & 245 & 2250 \\

$i$ & $90.26\pm0.13^\circ$ & $47.2^\circ$ & $77.2^\circ$ \\

$P_A$ (ms) & 22.7 & 59 & 37.9 \\

$\theta_A$ & $13^\circ\pm10^\circ$ & $21.1^\circ$ & $25.0^\circ\pm3.8^\circ$ \\

$\phi_A$ & $246^\circ\pm5^\circ$ & $9.7^\circ$ & $290^\circ\pm20^\circ$  \\

$P_{pA}$ (yr) & 74.9 & 297.2 & 700 \\


$\delta t_a/I_{A,80}~(\mu$s) & $0.17\pm0.16$ & 7.5  & $5.2\pm0.8$  \\

$A_{pA}/(A_{1PN} I_{A,80})$ & $6.6^{+0.2}_{-0.6}\times10^{-5}$ & $1.0
\times10^{-5}$ & $1.1 \times10^{-5}$ \\ 

$A_{2PN}/A_{1PN}$ & $5.1\times10^{-5}$ & $4.7\times10^{-5}$ & 
$2.3\times10^{-5}$ \\ \enddata

a: \cite{Lyne04}; b: Solution 1, \cite{Jenet04}; c: \cite{Coles04};

d: \cite{Weisberg02}; e: \cite{Weisberg04}; f: \cite{Stairs02}; 

g: \cite{Bogdanov02}; h: \cite{Stairs04b}

\end{deluxetable}

The precession of the spins and orbital plane of the pulsar binary PSR
J0737-3039 occurs with a period of about 75 years. Both the
inclination of the orbit along the line of sight and the position of
the periastron change due to spin-orbit coupling. The edge-on nature
of this binary, coupled with the slight misalignment of $\vec{S_A}$
and $\vec L$, probably precludes observation of the changing
inclination angle, but the spin-orbit contribution to the advance of
the perihelion, which amounts to a timing residual of nearly 0.03
seconds per year, should be measureable within a few years. In this
case, the edge-on orbital plane and near alignment of $\vec{S_A}$ and
$\vec L$ work in favor of a precision measurement of $I_A$. A measure
of $I_A$ to about 5-10\% accuracy seems possible,
and significant constraints on the equation of state would be
forthcoming. It would also lead to more reliable estimates of neutron
star radii and matter pressures near the nuclear saturation density
than are currently available.

Previously, the moment of inertia of the Crab pulsar was estimated by
\cite{Bejger03}. They used an estimate \citep{Fesen97} for the mass
of the ionized portion of the Crab's remnant, $4.6\pm1.8$ M$_\odot$,
to infer a lower limit to the Crab pulsar's moment of inertia of
$97\pm38$ M$_\odot$ km$^2$.  Within errors, the lower limit rules out
only the softest equations of state, as is shown in Fig. \ref{mom}.

It should be noted that the binary PSR J0737-3039, although highly
relativistic, is not favorably inclined for observation of the
precession of the inclination, which is proportional to $\cos i$. But
this particular inclination does allow a simplification of the
interpretation of the periastron advance.  Not only is the spin-orbit
contribution nearly independent of the pulsar orientation angle
$\phi_A$, but errors in $i$ are also not going to be a significant
restriction to a measurement of $I_A$.

It is interesting to compare PSR 0737-3039 with other known
relativistic binaries. Table 1 compares properties of the relativistic
binaries for which spin orientation of pulsar A can be estimated. It
is noteworthy that the net timing delays caused by precession-induced
inclination shifts are more than an order of magnitude larger for PSR
1913+16 and PSR 1534+12 than for PSR 0737-3039, due to their smaller
inclinations. However, the precessional periods of these two systems
are 4 times larger, and their periastron advances are about 6 times
smaller, than for PSR 0737-3039. This factor of 24 in observability is
significant, and explains why measurements of this effect in these
systems have not been made.
 
The nearness and faintness of this binary gives hope that other highly
relativistic systems might eventually be observed. If further highly
relativistic systems are discovered, it seems unlikely that any of them will
have an inclination angle so unfortunate for detecting the precession of the
orbital plane. With two spin-orbit effects to be observed, the moment of
inertia might be measured to even greater accuracy than contemplated for PSR
0737-3039.

\section*{ACKNOWLEDGEMENTS}

This work was supported in part by the U. S. Department of Energy 
under grant number DE-FG02-87ER-40317.   J.M.L. appreciates the hospitality
of the Albert-Einstein-Institut where this work was initiated.  Discussions
with C. Cutler are also greatly appreciated.

\end{document}